\pdfoutput=1
\documentclass{vgtc}

\ifpdf%
  \pdfoutput=1\relax                   %
  \usepackage{graphicx}                %
  \DeclareGraphicsExtensions{.pdf,.png,.jpg,.jpeg} %
\else%
  \usepackage{graphicx}                %
  \DeclareGraphicsExtensions{.eps}     %
\fi%

\graphicspath{{figures/}{pictures/}{images/}{./}} %
 
\usepackage{microtype}                 %
\PassOptionsToPackage{warn}{textcomp}  %
\usepackage{textcomp}                  %
\usepackage{mathptmx}                  %
\usepackage{times}                     %
\usepackage{cite}                      %
\usepackage{tabu}                      %
\usepackage{booktabs}                  %

\usepackage[final]{changes}
\usepackage{subcaption}

\definecolor{third}{HTML}{BAE4B3}
\definecolor{second}{HTML}{74C476}
\definecolor{first}{HTML}{238B45}
  
\usepackage{enumitem}
\setlist{nolistsep}
\setlist[itemize]{leftmargin=3mm}
\setlist[enumerate]{leftmargin=5mm}

\usepackage{titlesec}
\titlespacing*{\section}{0pt}{2.5ex plus 1ex minus 5ex}{1.3ex plus .2ex minus 3ex}

\usepackage{todonotes}

\usepackage{caption}

\onlineid{0}

\vgtccategory{Research}

\vgtcinsertpkg

\title{Interactive  Visualisation of Hierarchical Quantitative Data: An Evaluation}

\author{Linda Woodburn\thanks{e-mail: lwoo0004@student.monash.edu}\\ %
      \scriptsize Monash University
\and Yalong Yang\thanks{e-mail: yalong.yang@monash.edu}\\ %
      \scriptsize Monash University %
\and Kim Marriott\thanks{e-mail: kim.marriott@monash.edu}\\ %
      \scriptsize Monash University
}

\abstract{
We have compared three common visualisations for hierarchical quantitative data, treemaps, icicle plots and sunburst charts as well as a semicircular variant of sunburst charts we call the sundown chart. In a pilot study, we found that the sunburst chart was least preferred. In a controlled study with 12 participants, we compared treemaps, icicle plots and sundown charts. Treemap was the least preferred and had a slower performance on a basic navigation task and slower performance and accuracy in hierarchy understanding tasks. The icicle plot and sundown chart had similar performance with slight user preference for the icicle plot.  
} %

\CCScatlist{
  \CCScatTwelve{Human-centered computing}{Visu\-al\-iza\-tion}{Treemaps; Visualization design and evaluation methods}{}
}

\begin{document}
\maketitle

\section{Introduction}

Hierarchically organised quantitative data occurs in wide variety of contexts such as budget reports, CPU resource use, share portfolio allocation, and disk use. A number of different visualisations have been suggested to analyse such data~\cite{SchulzTreevisnetTree2011}. 
The best-known is the treemap~\cite{JohnsonTreeMapsSpacefillingApproach1991}. This is included in most modern data visualisation tools including Tableau and Microsoft Excel. Alternatives include the icicle plot,  recently used to show ``stack-traces''~\cite{GreggFlameGraph2016}, and  sunburst charts, used to visualise space allocation on a hard-disk~\cite{DaisyDiskDaisyDisk2018}.

However, to date there has been no comprehensive evaluation comparing the effectiveness of these three visualisations. 
Past evaluations~~\cite{Staskoevaluationspacefillinginformation2000,Barlowcomparison2Dvisualizations2001,PauloAlonsoGaonaGarciausabilitystudytaxonomy2014} have used small flat datasets, provided only very limited interactive navigation  and/or used low resolution displays.
The main contributions of this paper are:
\begin{itemize}
\item A novel visualisation, \emph{sundown chart}, a semicircular variant of sunburst chart that is better suited to the wider aspect ratio of modern displays and reduces user disorientation during zooming.
\item A carefully designed interactive navigation framework that is consistent across the four visualisations. This combines overview + detail, in-place zooming, breadcrumbs and uses unit visualisation~\cite{DruckerUnifyingFrameworkAnimated2015} to reduces user disorientation during zooming. 
\item A hybrid WebGL and D3 implementation that allows responsive interactive visualisation of large quantitative hierarchies with more than 50,000 nodes using standard internet browsing tools.
\item User evaluations of the visualisation designs. An initial pilot study with 6 participants compared all four visualisations. Feedback indicated that the sunburst chart was the least preferred. A second controlled study with 12 participants compared icicle plot, sundown chart and treemap. Treemap was clearly the worst, leading to slower performance task accuracy as well as being the least preferred. The icicle plot and sundown chart had similar performance with a slight user preference for the icicle plot.
\end{itemize}

\section{Related Work}  
\begin{figure*}[t!]
	\begin{subfigure}{\columnwidth}
	  \centering
	  \includegraphics[height=13.5em]{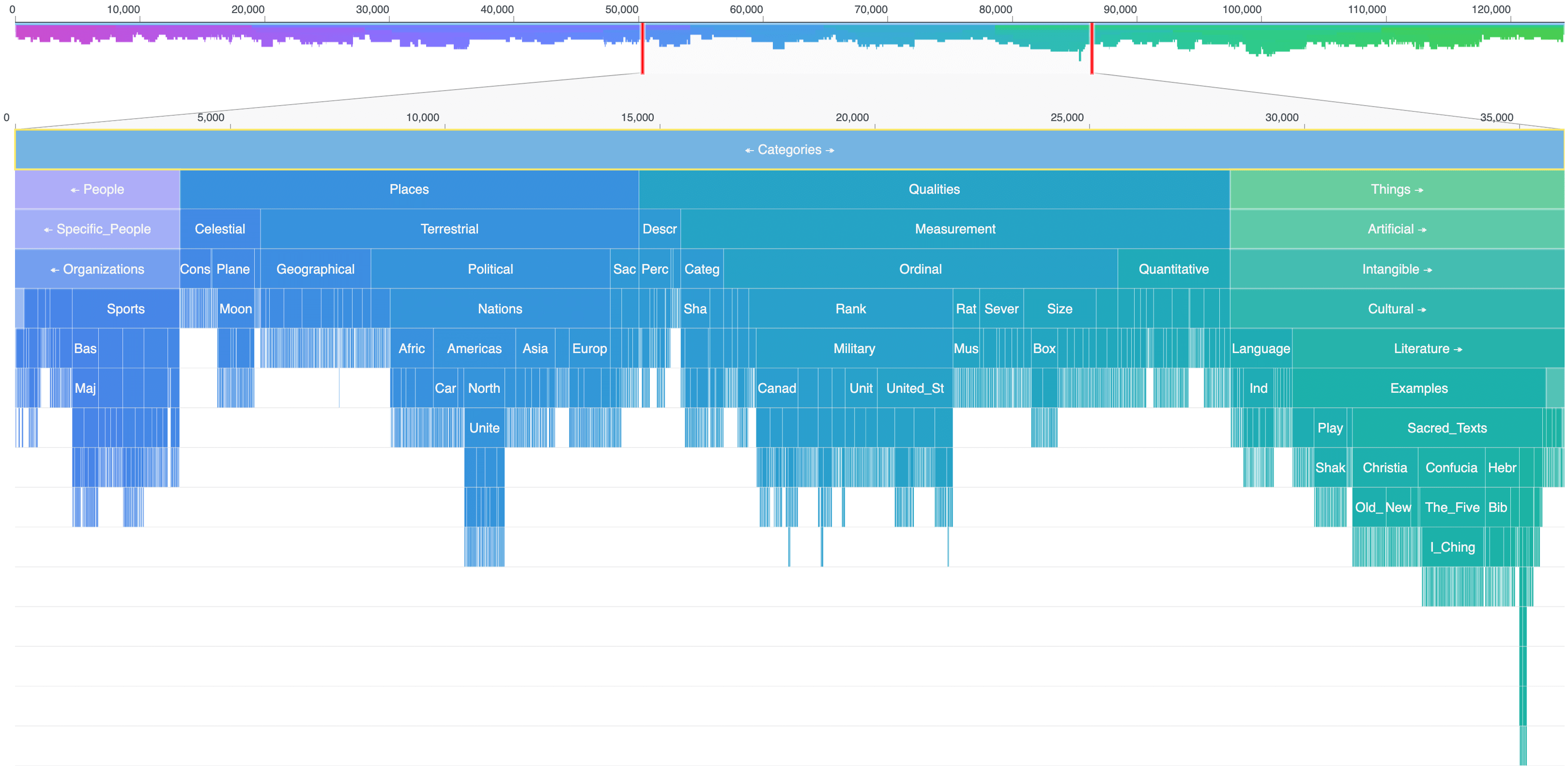}
	  \caption{Icicle plot}
	  \label{fig:finalvis:icicle}
	\end{subfigure}
	\begin{subfigure}{\columnwidth}
	  \centering
	  \includegraphics[height=13.5em]{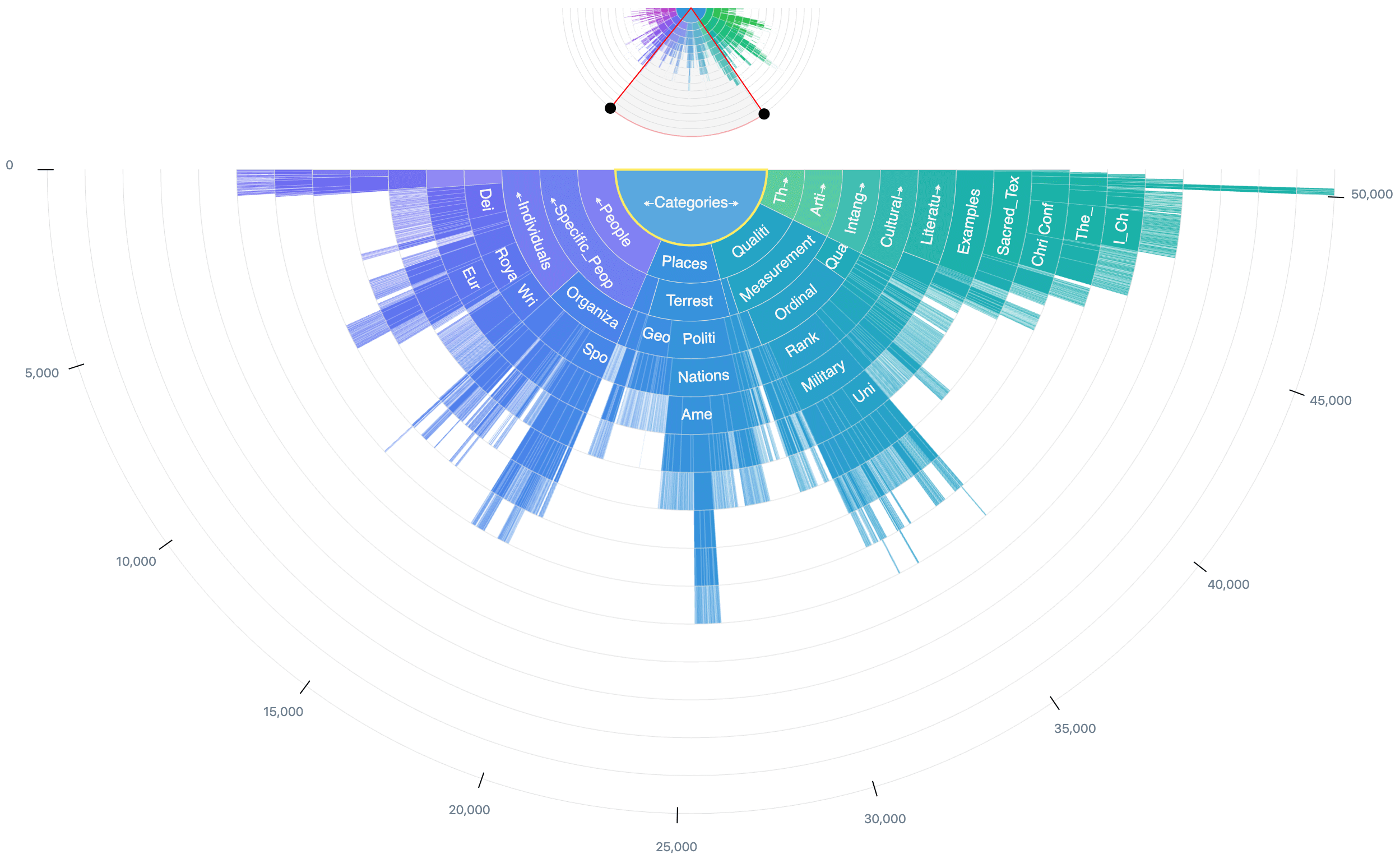}
	  \caption{Sundown chart}
	  \label{fig:finalvis:sundown}
	\end{subfigure}

	\begin{subfigure}{\columnwidth}
	  \centering
	  \includegraphics[height=13.5em]{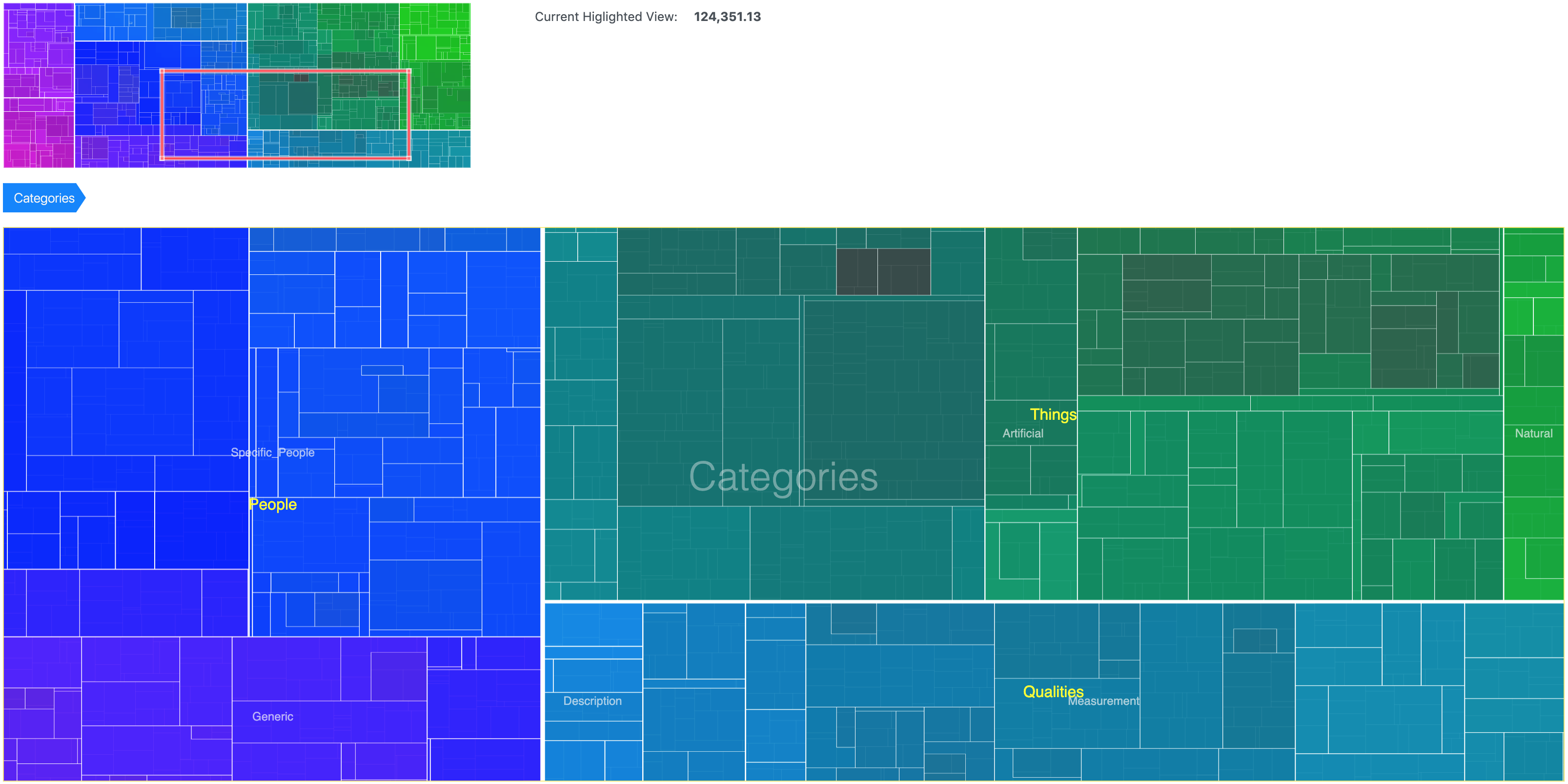}
	  \caption{Treemap}
	  \label{fig:finalvis:treemap}
	\end{subfigure}
	\begin{subfigure}{\columnwidth}
	  \centering
	  \includegraphics[height=13.5em]{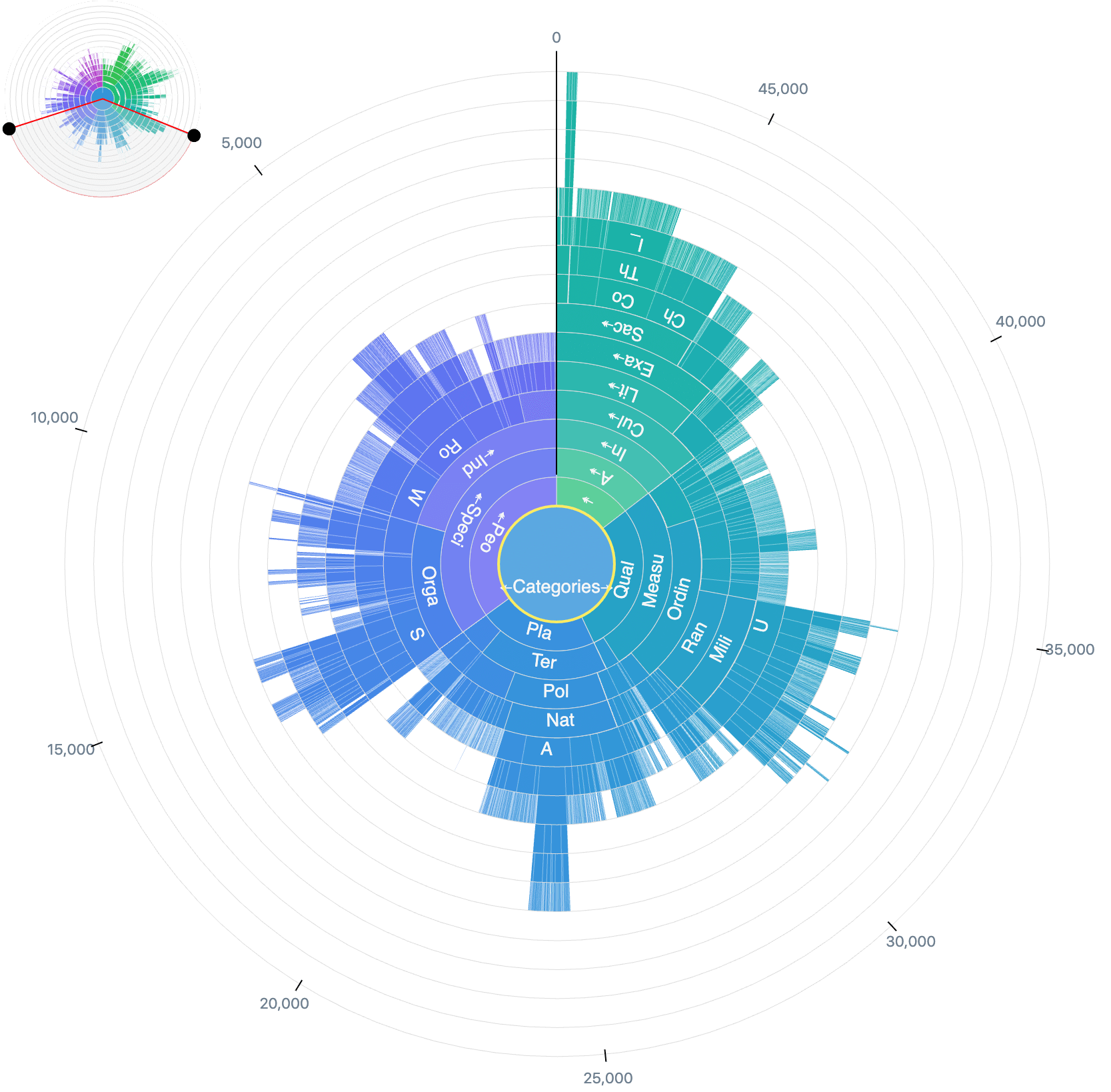}
	  \caption{Sunburst chart}
	  \label{fig:finalvis:sunburst}
	\end{subfigure}

  \caption{Final format of the visualisations for the user studies. All four provide an overview (top or top left) and a detailed view below. Treemaps also provide an explicit breadcrumb trail between the two views. All implementations and user study source code is open sourced at \url{https://github.com/ProcessFit/DataVis} and a demo working with \textit{Chrome} has been deployed at \url{https://qhd-vis.herokuapp.com/}.}
  \label{finalvis}
  \vspace{-1.5em}
\end{figure*}

\noindent\textbf{Treemaps}, see \autoref{finalvis}(c), use a space-filling design in which area denotes the quantitative node attribute and inclusion the node hierarchy~\cite{JohnsonTreeMapsSpacefillingApproach1991}. \added{They can be used to show both categorical and quantitative data: here we focus on quantitative data.} 
The original ``slice-and-dice'' algorithm~\cite{JohnsonTreeMapsSpacefillingApproach1991} led to nodes with high aspect ratio and unstable layout when the level of detail changed. Subsequent algorithms were designed to reduce aspect ratios~\cite{WattenbergVisualizingStockMarket1999,BrulsSquarifiedTreemaps2000} and to improve layout stability~\cite{BedersonOrderedQuantumTreemaps2003,SondagStableTreemapsLocal2018a}.

\noindent\textbf{Icicle plots}, see \autoref{finalvis}(a), were originally developed as a method for presenting hierarchical clustering~\cite{KruskalIciclePlotsBetter1983}. They use length to represent the quantitative attribute and juxtaposed layers to denote the hierarchy. Designs based on icicle plots have had a resurgence in recent years, e.g. SyncTRACE~\cite{KarranSYNCTRACEVisualthreadinterplay2013}, and Flame Graph~\cite{GreggFlameGraph2016}. 

\noindent\textbf{Sunburst charts}, see \autoref{finalvis}(d), are a radial version of the icicle plot. Each node is represented by a curved segment whose angle reflects the node quantitative attribute. An advantage of the radial layout  is the increased area to display nodes at deeper levels of the hierarchy. 
The ``Aggregate treemap''~\cite{ChuahDynamicaggregationcircular1998} and ``Information Slices''~\cite{AndrewsInformationSlicesVisualising1998} were early implementations of radial-space filling designs. The Aggregate treemap used an incomplete circle, and ``Information Slices'' used cascading semi-circular discs.
The original sunburst chart was introduced in~\cite{StaskoFocuscontextdisplay2000}. Like treemaps, the sunburst chart has achieved some popularity with recent integration into Microsoft Excel and  use in applications such as ``DaisyDisk''~\cite{DaisyDiskDaisyDisk2018}.

\noindent\textbf{Task taxonomies}
 developed for graphs and general visualisation techniques~\cite{LeeTaskTaxonomyGraph2006,SchulzDesignSpaceVisualization2013,Shneidermaneyeshaveit1996,AmarBESTPAPERKnowledge2004} provide a foundation for understanding the tasks specific to visualisations of hierarchical quantitative data. There are two main kinds of tasks:
understanding the \emph{hierarchy structure} (tree topology), e.g. depth or size of hierarchy, finding local relatives or distant relatives such as the least common ancestor, navigation through the hierarchy; 
and understanding the \emph{quantitative node attributes}, e.g. finding nodes with a specific value or extremums, comparing values.
In addition, there is the basic \emph{navigation} task of traversing the hierarchy.
It has been suggested that there is a conflict between designs that effectively represent hierarchies and those conveying the quantitative node attributes~\cite{SchulzDesignSpaceImplicit2011}.

\noindent\textbf{User evaluations:}
a number of studies have evaluated new more stable techniques for treemap layout ~\cite{YangCabinetTreeorthogonal2015,LuCascadedTreemapsExamining2008}.
Recommendations for improvements and best practice in the design of treemaps have been made, in terms of labelling, interactions~\cite{SedigDesignVisualizationsHumanInformation2016}, improvements to navigation~\cite{HornbaekNavigationPatternsUsability2003} and the provision of overviews~\cite{ElmqvistHierarchicalAggregationInformation2010}. 

Other researchers have produced design guidelines for creating perceptually effective treemaps, based on extensive user studies evaluating performance on value estimation tasks (comparison of node size)~\cite{KongPerceptualGuidelinesCreating2010}. The studies compared bar charts and treemaps, finding that treemaps become more effective than bar charts for size comparison tasks when the tree has thousands of leaf nodes. They recommended avoiding treemap layouts with extreme aspect ratios.

A few papers have compared treemaps and variants with Windows Explorer style visualisations~\cite{KobsaUserExperimentsTree2004,WangEvaluatingEffectivenessTree2006}.

We are aware of only three previous studies comparing the effectiveness of treemaps, sunburst charts, and icicle plots~\cite{Staskoevaluationspacefillinginformation2000,Barlowcomparison2Dvisualizations2001,PauloAlonsoGaonaGarciausabilitystudytaxonomy2014}.

One paper from 2000~\cite{Staskoevaluationspacefillinginformation2000} compared treemaps and sunburst charts. \added{It found a strong user preference for the sunburst chart and some indication that it had performance benefits for hierarchy understanding. However, little interaction was provided and the displays were low-resolution (800$\times$800)}. 

Another study from 2001 compared treemaps, sunburst charts, icicle plot and node-link diagrams~\cite{Barlowcomparison2Dvisualizations2001}. Only hierarchy structure understanding tasks were considered. The treemap was slower and less accurate and the icicle plot and node-link were preferred. However, the implementations provided limited functionality for navigation, \added{were limited by computational power and screen resolution (1024$\times$768) and used a small dataset ($<$200 nodes) in the evaluation}.

More recently, a study compared the usability of various visualisations including treemaps, icicle plots and sunburst charts for search and exploration of large digital collections~\cite{PauloAlonsoGaonaGarciausabilitystudytaxonomy2014}. A JavaScript InfoVis Toolkit~\cite{JavaScriptInfoVisToolkit2013} providing a common platform for the visualisations used. This was limited in terms of navigation and provided no overview of the structure.
The results indicated that icicle plots and tree designs were more effective for visualising trees with greater depths. It was also noted that the  dataset being visualised caused computational issues ``compromising the performance of the platform.'' Tasks relating to the identification and comparison of the quantitative attributes of the underlying data were not performed. 

\added{Interactive \textbf{navigation} of larger hierarchical datasets has been considered by a number of researchers. Most commonly, users can collapse or expand nodes~\cite{plaisant2002spacetree} and/or zoom to lower levels of the hierarchy using \emph{zoom-and-replace}, where the view associated with the selected node replaces that of its parent~\cite{BlanchBrowsingZoomableTreemaps2007, ChintalapaniExtendingutilitytreemaps2004,DaisyDiskDaisyDisk2018}. We believe this behaviour and the lack of overview can lead to issues with navigation. Flame Graph~\cite{GreggFlameGraph2016} uses \emph{zoom-in-place} in which the selected node expands to fill the viewport but does not provide an overview. SyncTRACE~\cite{KarranSYNCTRACEVisualthreadinterplay2013} also uses zoom-in-place but provides a two-level overview. Information Slices uses linked cascaded views at increasing levels of detail to provide contextual information~\cite{AndrewsInformationSlicesVisualising1998} while  two-level views with different spatial arrangements have been suggested in ~\cite{StaskoFocuscontextdisplay2000}. However, none of these designs have been used in user studies comparing different visual representations for quantitative hierarchical data.}

Given the importance of hierarchical quantitative data and the fact that icicle plots and sunburst charts are increasingly used in practical applications, it is timely to conduct another comparison with treemaps, the most widely used visualisation for hierarchical quantitative data, using larger datasets and providing more comprehensive support for navigation in such larger datasets.

\vspace{1em}
\section{Visualisations}
\vspace{0.5em}

In order to fairly compare the different visualisations, we implemented them using a consistent design framework. This framework was designed to overcome the previously identified difficulties of navigation within larger datasets.

\added{\noindent\textbf{Sundown Chart:}
as part of the study we also introduced a hybrid of the icicle plot and sunburst charts that we called the \emph{sundown chart}, see \autoref{finalvis}(b). The sundown chart is a semi-circular design, with the root node placed at the centre of the design, and child nodes fanning out from the root node. The design is a rotated version of a single Information Slice~\cite{AndrewsInformationSlicesVisualising1998} and is similar to a ``fan chart'' that is often used to represent genealogical hierarchies~\cite{draper2008interactive}, with the additional encoding of the quantitative variable in a similar manner to the sunburst chart.}

The sundown chart design provides the benefits of both the icicle plot and the sunburst chart. Similarly to the icicle plot, the aspect ratio is better suited to the 16:9 aspect ratio of a computer monitor, when compared to the 1:1 aspect ratio of the sunburst chart. Like the sunburst chart, the radial design provides proportionally more space for the smaller nodes at lower level of the hierarchies. But compared to the sunburst chart labels are orientated more consistently.

\noindent\textbf{Navigation:}
we chose to provide an \emph{Overview + Detail} view of the hierarchy. For each visualisation, a compact overview of the hierarchy was shown above the main detailed view. For the treemap, both height and width were reduced in the overview while for the other visualisations only the non-quantitative dimension.

Like previous quantitative hierarchical data visualisations, we allowed the user to navigate the hierarchy using \emph{zoom and pan} and \emph{drill-down} from parent to child node and \emph{roll up} to return. \emph{Zoom and pan} is either from the detail view, or by using the ``handles'' shown in the overview to manipulate the position of the detail view.
\added{Inspired by Flame Graph~\cite{GreggFlameGraph2016} and SyncTRACE~\cite{KarranSYNCTRACEVisualthreadinterplay2013}  we chose to use \emph{zoom-in-place} which increases the zoom in the same dimensions as the node dimensions as this preserves more contextual information.} For example, zoom actions in icicle plots increase the zoom on the x-axis only, increasing the  width of the node. For sunburst charts, where the node size is proportional to the angle of the arc segments, the angle is proportionally increased. For treemaps, zoom takes place in both the x and y axes, proportional to the area of the node.

In order to support zoom-in-place, we chose to avoid aggregation or elision of the nodes in either of the views, and instead showed all nodes. This was inspired by ``Interactive information visualization of a million items''~\cite{FeketeInteractiveInformationVisualization2003} and ``Unit Visualisations''~\cite{DruckerUnifyingFrameworkAnimated2015}. This allowed us to smoothly animate zoom-in-place transitions and ensured layout stability, as all of the nodes are already present.

An advantage of our zoom-in-place design is that, in the icicle plot, sundown and sunburst charts, higher levels of the hierarchy are not removed from the visualisation when the user selects a lower-level node. This results in a stack of ``breadcrumbs'' remaining visible as the user zooms to lower level nodes, showing the path between the root node and the currently selected node. In order to level the playing field we also provided an explicit breadcrumb trail for the treemap design. 

We used a dynamic ``label what you can'' approach. Labels are centred on the visible portion of the nodes, so that during interaction labels remain visible even as the underlying node is panned outside of the current viewpoint. Where partial nodes are shown, the label indicated the ``off-screen'' portion of the node with chevrons. Tooltip display of labels was also provided.

The user was provided with the ability to tag nodes as this was required for the study. Tagged nodes appear in the overview as ``tag-dots'' and as node outlines in the detail view.

For all designs, we included a scale to assist in size comparison tasks and to clearly show the magnification factor in the detail view. %

\noindent\textbf{Other Design Choices:}
\added{following InterRing~\cite{YangInterRinginteractivetool2002}, we used a structure-based colouring strategy to colour nodes along a ``cubehelix'' spectrum~\cite{green2011colour} in all visualisations. Hue discriminates between different branches and emphasizes the hierarchical structure. Other colouring strategies remain to be explored, \textit{e.g.}~\cite{tennekes2014tree}.}

We used a squarified treemap layout, based on the guidelines produces by Kong~\textit{et al.}~\cite{KongPerceptualGuidelinesCreating2010}. %
To further emphasize hierarchical nesting we used relative colour saturation, with deeper nodes appearing more saturated (similar to the colours implemented by Rosenbaum and Hamann~\cite{RosenbaumEvaluationprogressivetreemaps2012}). 

\noindent\textbf{Implementation:}
initially, the visualisations were implemented using the popular D3 JavaScript library. However we found that the visualisations created using D3 were limited to about 800-1000 elements, beyond which the animations become unresponsive.

We then explored WebGL. This utilises the client machine's GPU (Graphical Processing Unit), and is able to render significantly more elements than D3 can while supporting responsive animation and interaction. 
The final designs were implemented using a hybrid WebGL and D3 model, in which the nodes and visual elements were implemented using a WebGL layer, with labels superimposed using D3. Using this hybrid model, we were able to produce highly dynamic and smooth visualisations that do not need to aggregate or elide nodes with visualisations containing in excess of 50,000 nodes, with rendering rates at 50-60 frames per second.

\section{Evaluation}
We evaluated our visualisations first in a pilot study then a larger controlled study.

\subsection{Pilot Study}
A pilot study was conducted with six participants. Five were male, one was female. All were post-graduate data science students with some experience in data visualisation. Three participants were in the age range 18-25, two in the age range 25-35 and one over the age of 45 years. They were shown each of the four visualisations and asked general questions relating to the aesthetics of the visualisation, then to perform a range of tasks that require interacting with the visualisations. The study took approximately 90 minutes.

Feedback from the participants was used to improve the visualisation designs and to refine the tasks used in the full user study.

We used the \textit{CHI Browse-Off} categories hierarchy dataset~\cite{mullet1997your}. This has 15 levels, 7,548 nodes of which 6,278 are leaf nodes. 

All visualisations apart from the sunburst chart received positive feedback. That for the sunburst was mixed. While the initial impression of the radial design were positive, participants found that the visualisation was more difficult to interpret. The relatively large amount of white space required by the full 360\textdegree~design with a 1:1 aspect ratio resulted in smaller sized nodes overall, and subsequently less labeling of nodes. Additionally, techniques used to zoom-in-place lead to ``disappearing'' nodes along one radial axis (\textit{i.e.} for zoomed views, partial nodes are shown on both sides of the 12:00 position) that was perceptually more difficult to understand than partial nodes displayed at the ``edges'' of the design seen in the sundown chart and icicle plot designs.
Based on this negative feedback, the sunburst chart was excluded from the controlled user study to allow more time for the evaluation of the other designs.

\subsection{Controlled User Study}
We then conducted a larger controlled user study to compare the effectiveness of the three visualisations in terms of speed, errors and user preference. The visualisations, including an overview, scales and other design elements are shown in \autoref{finalvis}.

\noindent
\textbf{Participants:}  
Twelve participants were recruited through a closed-group on social media for Monash information technology students and staff. None had taken part in the pilot study. While most of the participants had seen treemaps previously, only two had seen icicle plots. No participants had previously seen sundown charts. Eight participants were in the age range 18-25, one in the age range 25-35 and three over the age of 45 years. Eight were male, four female. 

\noindent
\textbf{Tasks:}
Four tasks were evaluated. These checked basic navigation (Q1),
understanding of quantitative node attributes (Q2) and hierarchy structure (Q3, Q4):
\begin{itemize}
    \item \textbf{Q1} - ``Navigate to a specific node'' This task required  the participant to navigate along a explicitly given path of the hierarchy to reach a leaf node, and then ``tag'' the node. 

    \item \textbf{Q2} - ``Compare the size of two nodes'' This required the participant to tag the larger of two pre-tagged nodes. The size difference between the two nodes was between 10 and 40\%. 

    \item \textbf{Q3} - ``Identify the closest relationship between Node A and Node B'' Participants were asked to choose the relationship that most precisely described that between two tagged nodes:
    sibling, parent, child, ancestor.
    
    \item \textbf{Q4} - ``What is the least common ancestor of Node A and Node B'' Participants were asked to tag the least common ancestor of two tagged nodes.
\end{itemize}

\noindent
\textbf{Procedure:}
The user study was in the following process:  
\begin{itemize}
 \item Participants were welcomed, introduced to the study, and had the evaluations process explained.
 \item For each visualisation, \emph{training} was provided and then twelve \textbf{trial tasks} were presented to the participant, three items for each of the four task types.
 \item \added{Following the completion of the evaluations tasks for all visualisation, the user was asked to answer a post-study questionnaire: 1) How would you rank the visualisations in terms of aesthetics, \textit{i.e.} visually pleasing, engaging? 2) How would you rank the visualisations in order of usability, \textit{e.g.} ability to navigate and interact with the visualisation?  3) Which visualisation do you prefer overall?}
\end{itemize}

Presentation order of visualisations was counterbalanced.
Training was delivered to participants using an in-application interactive tutorial. The training presented the participant with background information on quantitative hierarchical data, a description and example of the structure of each visualisation, as well as instructions on how to use the available interactions. Sample visualisations  were used to confirm the participant's understanding of the structure and attributes of each visualisation. Finally, practice questions were provided to the participant, along with feedback relating to the accuracy of the answer. 

\added{We used a dataset based on the \textit{CHI Browse-Off} dataset, where each leaf node was assigned a random quantitative value. In the original dataset, all leaf nodes had the same value and we wished to increase the complexity of the data.}
3 sets of 12 questions were randomly generated and hand-curated to ensure similar level of difficulty. Presentation of these question sets were counterbalanced between the three visualisations.

All users were tested using the same computer set-up, consisting of a Mac computer with a 26-inch monitor (a resolution of 2560$\times$1440), and a standard mouse equipped with a scroll wheel.

The study took about 1 hour to complete.

\noindent
\textbf{Measures:}
The time taken to complete each task (from initial rendering to answer selection), accuracy (binary--correct/incorrect), and mouse-interactions %
were logged by the application.

\vspace{-0.5em}
\subsection{Results and Discussion}
Answers were not normally distributed. Significance was tested with the Friedman test because we have more than two conditions and a post-hoc test was used to compare pairwise. Response times were log-normal distributed (checked with histograms and Q---Q plots), so a log-transform was used before statistical analysis~\cite{howell2012statistical}. We chose linear mixed modeling to check for significance~\cite{bates2015fitting} and applied Tukey's HSD post-hoc tests to conduct pairwise comparisons~\cite[p.~575-582]{field2012discovering}. A within-subject design with random intercepts were used for all models. For user preferences we again used the Friedman test and post-hoc test to test for significance.

\begin{figure}[t]
\centering
\includegraphics[width=0.95\columnwidth]{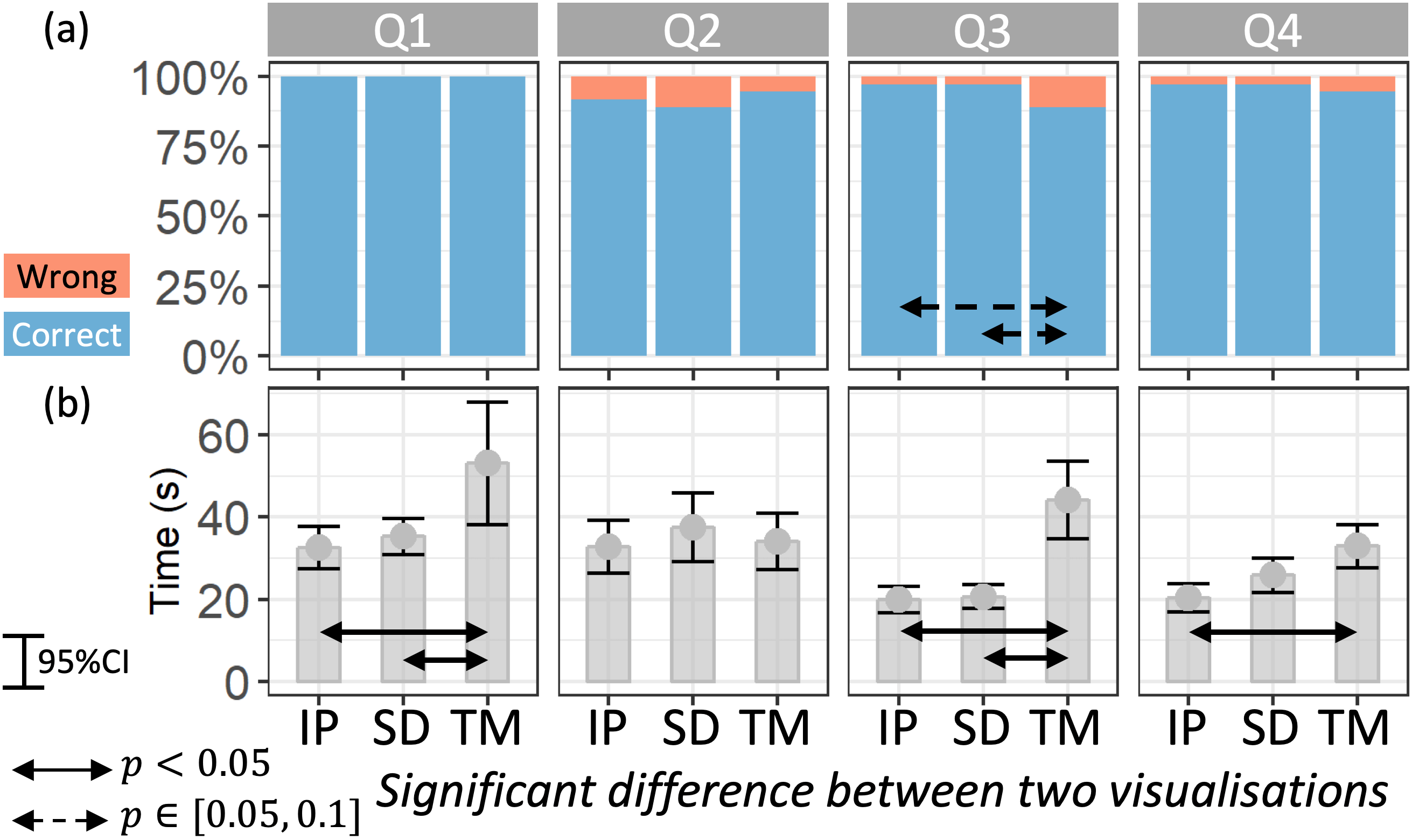}
\caption{Visualisation performance for different tasks: (a) percentage of correct and wrong responses; (b) 95\% confidence interval of time. IP---icicle plot, SD---sundown chart and TM---treemap.}
\label{fig-results}
\vspace{-2em}
\end{figure}

The Friedman test revealed no significant effect of visualisations on accuracy for Q1, Q2 and Q4, but a significant effect for Q3 ($\chi^2(2)=6, p=.0498$). \autoref{fig-results} (a-Q3) shows icicle plot (97.2\%) and sundown chart (97.2\%) were more accurate than  treemap (91.4\%). Post-hoc tests showed for both $p=.0855$.

The linear mixed modeling analysis showed significant effects of visualisations on time for Q1, Q3 and Q4. 

\noindent\textbf{Q1:} $\chi^2(2)=17.47, p=.0002$. Icicle plot (avg. 32.6s) and sundown chart (avg. 35.2s) were significant faster than treemap (avg. 53.0s). All $p<.05$.

\noindent\textbf{Q3:} $\chi^2(2)=23.95, p<.0001$. Icicle plot (avg. 20.0s) and sundown chart (avg. 20.6s) were significant faster than treemap (avg. 44.2s). All $p<.0001$.

\noindent\textbf{Q4:} $\chi^2(2)=11.18, p=.0037$. 
Icicle plot (avg. 20.3s) was significant faster than treemap (avg. 33.0s) with $p=.0005$. While sundown chart (avg. 25.8s) tended to be faster than treemap, see \autoref{fig-results} (b-Q4), it was not statistically significant.
  
The three visualisations had similar performance for Q2. We had expected that efficiency and accuracy would be greatest for the icicle plots and least for the treemap based on previous work~\cite{ClevelandGraphicalPerceptionTheory1984}. It suggested that size comparisons for one-dimensions (line length) are easier than angular comparisons, which are in turn easier than comparisons in two-dimensions (area). The lack of difference may because of that all three visualisations provided a scale. 
  
\begin{figure}[t]
\centering
\includegraphics[width=0.8\columnwidth]{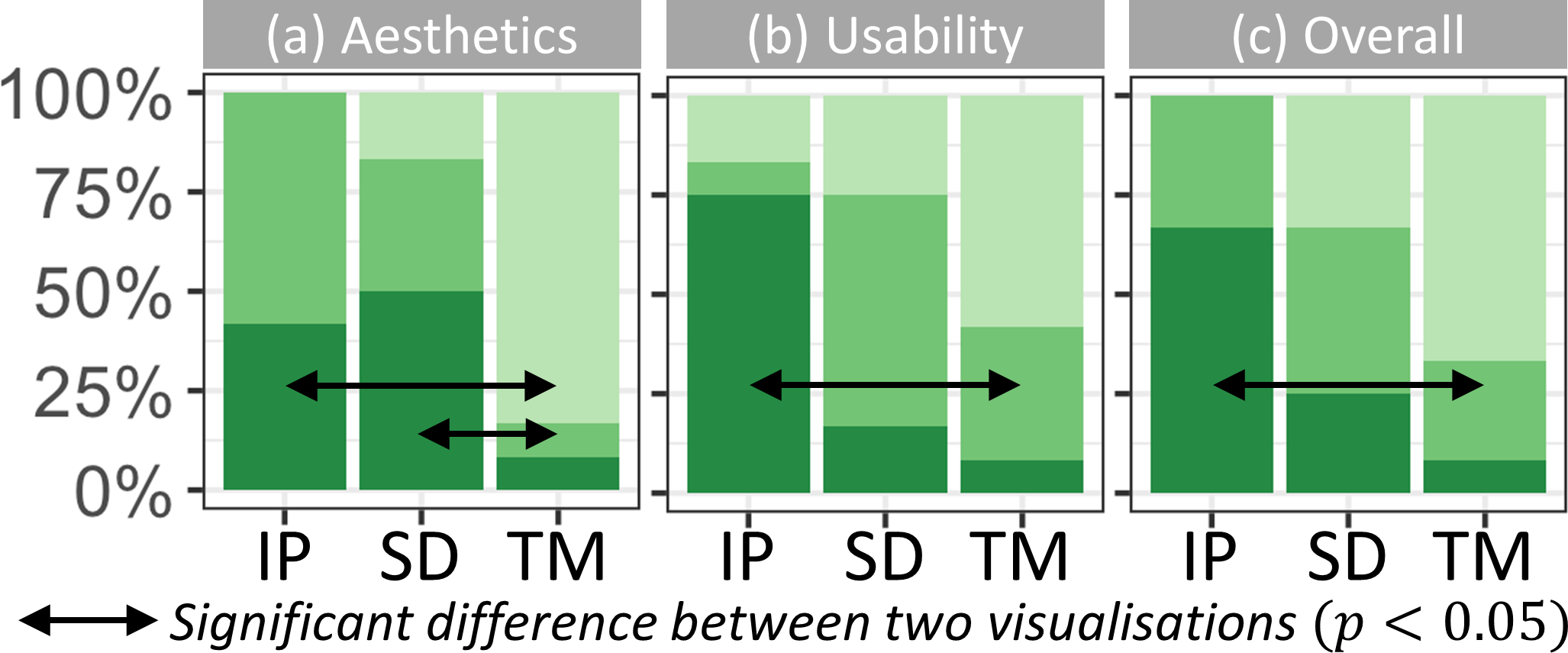}
\caption{Participants preference ranking (\setlength{\fboxsep}{1.5pt}\colorbox{first}{\textcolor{white}{$1^{st}$}}, \setlength{\fboxsep}{1pt}\colorbox{second}{\textcolor{white}{$2^{nd}$}} and \setlength{\fboxsep}{1pt}\colorbox{third}{$3^{rd}$}). IP---icicle plot, SD---sundown chart and TM---treemap.}
\label{fig-rankings}
\vspace{-1.5em} 
\end{figure}

\noindent\textbf{User preference and feedback:} 
For \emph{aesthetics}, see \autoref{fig-rankings}(a), the Friedman test revealed a significant effect of visualisations on preference ($\chi^2(2)=11.54, p=.0031$). Participants liked the visual design of icicle plot 
and sundown chart 
significantly more than treemap. 
All $p<.05$.

For \emph{usability}, see \autoref{fig-rankings}(b), the Friedman test revealed a significant effect of visualisations on preference ($\chi^2(2)=7.17, p=.0278$). Participants preferred the usability of icicle plot 
significantly more than treemap 
with $p=.0218$. Icicle plot was tended to be more preferred than sundown chart, 
however this was not significant.

\emph{Overall}, see \autoref{fig-rankings}(c), the Friedman test revealed a significant effect of visualisations on preference ($\chi^2(2)=9.5, p=.0087$). Participants preferred the icicle plot overall, 
significantly more than treemap 
with $p=.0218$. Icicle plot was also tended to be  preferred to the sundown chart, 
however this was not statistically significant.

The final section of the study allowed participants to give feedback on the pros and cons of each visualisation as well as the overall design. Qualitative analysis of these comments reveal (overall):

\noindent\textbf{\textit{Icicle plot}} was found to be intuitive and easy to learn. Some participants also mentioned it uses the screen space more effectively than a sundown chart but less than a treemap.

\noindent\textbf{\textit{Sundown chart}} was commented as novel. However, we observed almost all participants tended to occasionally rotate their heads  and some participants reported feeling slight motion sickness.

\noindent\textbf{\textit{Treemap}} was found difficult to start with, but \textit{``it got better''}. Some participants found breadcrumbs useful. Many participants had difficulties navigating the hierarchy.

\section{Conclusion} 
We have compared three common visualisations for hierarchical quantitative data, treemaps, icicle plots and sunburst charts as well as a semicircular variant of sunburst charts we call the sundown chart. In a pilot study we found that the sunburst chart was least preferred. In a controlled study with 12 participants we compared the remaining three visualisations. Overall treemap was clearly the worst, leading to slower performance on a basic navigation task and slower performance and accuracy in hierarchy understanding tasks. It was also the least preferred by users. The icicle plot and sundown chart had similar performance with a slight user preference for the icicle plot.
Interestingly we found no evidence of the conflict suggested by Schulz~\cite{SchulzDesignSpaceImplicit2011} between designs that effectively represent hierarchies and those conveying the quantitative node attributes: icicle plots and sundown charts appear to work well for both tasks.
  
\added{A limitation of the current study is the use of randomised quantitative data in the user studies which may be unrepresentative of real-world data sets. Additional studies are required to confirm the findings for other datasets including different shaped hierarchies.}
 
\acknowledgments{This work was supported by the  Australian Research Council Discovery Project grant DP140100077.}

\bibliographystyle{abbrv}

\bibliography{template}
\end{document}